\documentclass[12pt]{article}
\usepackage{graphicx,amsmath,amssymb,cite}

\newlength{\dinwidth}
\newlength{\dinmargin}
\newcommand{\dif}{\mathrm{d}}
\newcommand{\diff}[1]{\frac{\mathrm{d}#1}{#1}}
\newcommand{\xB}{x_{\scriptscriptstyle{B}}}
\newcommand{\Pom}{{\hspace{ -0.1em}I\hspace{-0.2em}P}}
\newcommand{\xPom}{x_\Pom}
\newcommand{\chisq}{\chi^2/\mathrm{d.o.f.}}

\begin{document}
\titlepage
\begin{flushright}
  DESY 05-136      \\
  IPPP/05/41       \\
  DCPT/05/82       \\
  8th August 2005  \\
\end{flushright}

\vspace*{0.5cm}

\begin{center}
  
  {\Large \bf Effect of absorptive corrections \\[1ex] on inclusive parton distributions}

  \vspace*{1cm}

  \textsc{G. Watt$^a$, A.D. Martin$^b$ and M.G. Ryskin$^{b,c}$} \\

  \vspace*{0.5cm}

  $^a$ Deutsches Elektronen-Synchrotron DESY, 22607 Hamburg, Germany \\
  $^b$ Institute for Particle Physics Phenomenology, University of Durham, DH1 3LE, UK \\
  $^c$ Petersburg Nuclear Physics Institute, Gatchina, St.~Petersburg, 188300, Russia

\end{center}

\vspace*{0.5cm}

\begin{abstract}
  We study the effect of absorptive corrections due to parton recombination on the parton distributions of the proton.  A more precise version of the GLRMQ equations, which account for non-linear corrections to DGLAP evolution, is derived.  An analysis of HERA $F_2$ data shows that the small-$x$ gluon distribution is enhanced at low scales when the absorptive effects are included, such that a negative gluon distribution at 1 GeV is no longer required.
\end{abstract}

\section{Introduction}

At very small values of $x$ it is expected that the number density of partons within the proton becomes so large that they begin to recombine with each other.  This phenomenon of parton recombination is also referred to as absorptive corrections, non-linear effects, screening, shadowing, or unitarity corrections, all leading to saturation.  The first perturbative QCD (pQCD) calculations describing the fusion of two Pomeron ladders into one were made by Gribov-Levin-Ryskin (GLR) \cite{Gribov:1984tu} and by Mueller-Qiu (MQ) \cite{Mueller:1985wy}.  The GLRMQ equations add an extra non-linear term, quadratic in the gluon density, to the usual DGLAP equations for the gluon and sea-quark evolution.  The evolution of the gluon distribution is then given by
\begin{equation} \label{eq:GLRMQgluon}
  \frac{\partial xg(x,Q^2)}{\partial\ln Q^2} = \frac{\alpha_S}{2\pi}\sum_{a^\prime=q,g}P_{ga^\prime}\otimes{a^\prime} ~-~ \frac{9}{2}\,\frac{\alpha_S^2(Q^2)}{R^2\,Q^2}\,\int_x^1\!\diff{x^\prime}\left[x^\prime g(x^\prime,Q^2)\right]^2,
\end{equation}
where $R\sim 1$ fm is of the order of the proton radius.  The GLRMQ equations account for all `fan' diagrams, that is, all possible $2\to1$ ladder recombinations, in the double leading logarithmic approximation (DLLA) which resums all powers of the parameter $\alpha_S\ln(1/x)\ln(Q^2/Q_0^2)$.

There has been much recent theoretical activity in deriving (and studying) more precise non-linear evolution equations, such as the Balitsky-Kovchegov (BK) and Jalilian-Marian--Iancu--McLerran--Weigert--Leonidov--Kovner (JIMWLK) equations (see \cite{Jalilian-Marian:2005jf} for a review).  Note that the BK and JIMWLK equations are both based on BFKL evolution.  However, for the most relevant studies in the HERA and LHC domain ($x\gtrsim 10^{-4}$), the predominant theoretical framework is collinear factorisation with DGLAP-evolved parton distribution functions (PDFs).  At very small values of $x$ it might be expected that the DGLAP approximation would break down, since large $\alpha_S\ln(1/x)$ (BFKL) terms would appear in the perturbation series in addition to the $\alpha_S\ln(Q^2/Q_0^2)$ terms resummed by DGLAP evolution.  However, it turns out that the resummed NLL BFKL calculations of the gluon splitting function $P_{gg}$ \cite{smallxpgg} and the gluon transverse momentum distribution \cite{Khoze:2004hx} are rather close to the DGLAP calculations.  Moreover, the convolution $P_{gg}\otimes g(x,Q^2)$ coincides with the NNLO DGLAP result and is close to the NLO DGLAP result for $x\gtrsim 10^{-4}$ \cite{Salam:2005yp}.  Hence, in the analysis of current data, it is reasonable to ignore BFKL effects.

If recombination effects are significant, it is therefore important that they be incorporated into the global DGLAP parton analyses which determine the PDFs from deep-inelastic scattering (DIS) and related hard-scattering data.  Such a programme, based on GLRMQ evolution (which accounts for gluon-induced screening only), was implemented some years ago \cite{Kwiecinski:1990ru}, before the advent of HERA.  The input gluon and sea-quark distributions were \emph{assumed} to have a small-$x$ behaviour of the form $xg,xS\sim x^{-0.5}$ at an input scale of $Q_0^2=4$ GeV$^2$.  The inclusion of shadowing effects, both in the form of the input PDFs and in the GLRMQ evolution, was found to significantly \emph{decrease} the size of the small-$x$ gluon distribution in comparison with the result with no absorptive corrections.  A crucial observation is that, at that time (1990), $F_2$ data were only available for $\xB\ge 0.07$, and so these results were largely dependent on the theoretical assumptions made for the starting distributions.  However, with HERA, we now have $F_2$ data down to $\xB\sim 10^{-4}$ or less, and so the PDFs at small $x$ can be determined directly from the HERA data.

\begin{figure}
  \centering
  \includegraphics[width=0.5\textwidth]{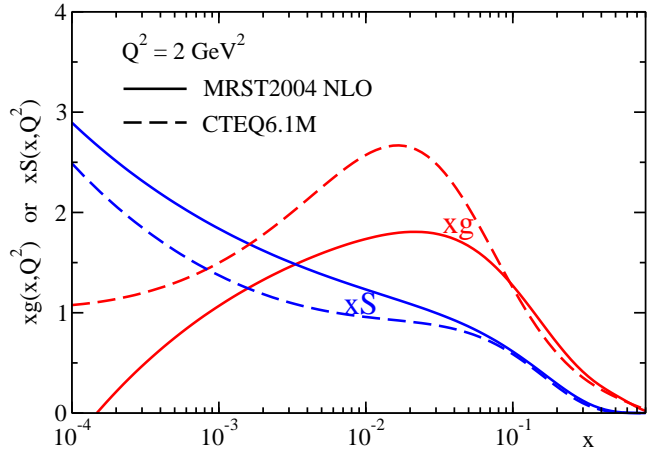}
  \caption{The behaviour of the gluon and sea-quark distributions at $Q^2=2$ GeV$^2$ found in the MRST2004 NLO \cite{Martin:2004ir} and CTEQ6.1M \cite{Pumplin:2002vw} global analyses.  The valence-like behaviour of the gluon is evident.}
  \label{fig:mrstcteq}
\end{figure}

In fact, the advent of HERA data has led to a puzzling behaviour of the small-$x$ gluon and sea-quark PDFs at low scales $Q^2$.  If we write $xg\sim x^{-\lambda_g}$ and $xS\sim x^{-\lambda_S}$, then the expectation of Regge theory is that $\lambda_g=\lambda_S=\lambda_{\text{soft}}$ for low scales $Q\lesssim Q_0\sim 1$ GeV, where $\lambda_{\text{soft}}\simeq 0.08$ \cite{Donnachie:1992ny} is the power of $s$ obtained from fitting soft hadron data.  At higher $Q\gtrsim 1$ GeV, QCD evolution should take over, increasing the powers $\lambda_g$ and $\lambda_S$.  However, the current MRST2004 NLO \cite{Martin:2004ir} and CTEQ6.1M \cite{Pumplin:2002vw} PDF sets exhibit a very different behaviour at low scales from that theoretically expected; see Fig.~\ref{fig:mrstcteq}.  In fact, the MRST group has found that a negative input gluon distribution at $Q_0 = 1$ GeV is required in all their NLO DGLAP fits since MRST2001 \cite{Martin:2001es}.  The CTEQ group, who take a slightly higher input scale of $Q_0 = 1.3$ GeV, also find a negative gluon distribution when evolving backwards to 1 GeV.

Since data at small $\xB$ now exist, the introduction of the absorptive corrections is expected to \emph{increase} the size of the input gluon distribution at small $x$ to maintain a satisfactory fit to the data.  To understand this, note that the negative non-linear term in the GLRMQ equation \eqref{eq:GLRMQgluon} slows down the evolution.  Therefore, it is necessary to start with a \emph{larger} small-$x$ gluon distribution at low scales $Q\sim Q_0$ to achieve the \emph{same} PDFs at larger scales required to describe the data.  If the non-linear term is neglected, the input small-$x$ gluon distribution is forced to be artificially small in order to \emph{mimic} the neglected screening corrections.

We have anticipated that the introduction of absorptive corrections will \emph{enhance}\footnote{Eskola \emph{et al.}~\cite{Eskola:2002yc} have found that taking input gluon and sea-quark distributions at $Q^2=1.4$ GeV$^2$, then evolving upwards with the GLRMQ equations based on LO DGLAP evolution, improves the agreement with $F_2$ data at small $\xB$ and low $Q^2$ compared to the standard CTEQ sets, and leads to an enhanced small-$x$ gluon distribution for $Q^2\lesssim 10$ GeV$^2$.  Note, however, that there is a large NLO correction to the splitting function $P_{qg}$ which changes completely the relationship between the quark and gluon distributions, and so weakens the conclusion of Ref.~\cite{Eskola:2002yc}.} the small-$x$ gluon at low scales, and hence could possibly avoid what appears to be anomalous behaviour at small $x$.  Thus, here, we perform such a study using an abridged version of the MRST2001 NLO analysis \cite{Martin:2001es}, improving on our previous analysis \cite{Martin:2004xx}.  First, we derive a more precise form of the GLRMQ equations.

\section{Non-linear evolution from diffractive DIS} \label{sec:non-linear}

\begin{figure}
  \centering
  \begin{minipage}{0.8\textwidth}
    (a)\\
    \includegraphics[width=\textwidth]{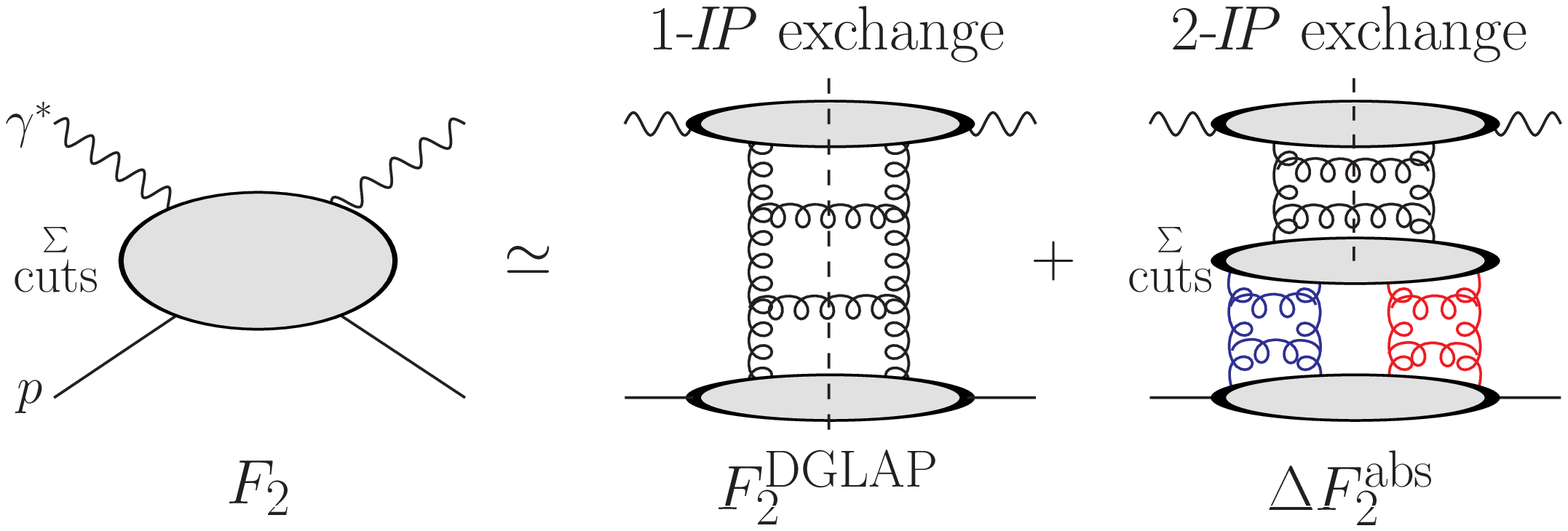}
  \end{minipage}\\[2mm]
  \begin{minipage}{0.8\textwidth}
    (b)\\
    \includegraphics[width=\textwidth]{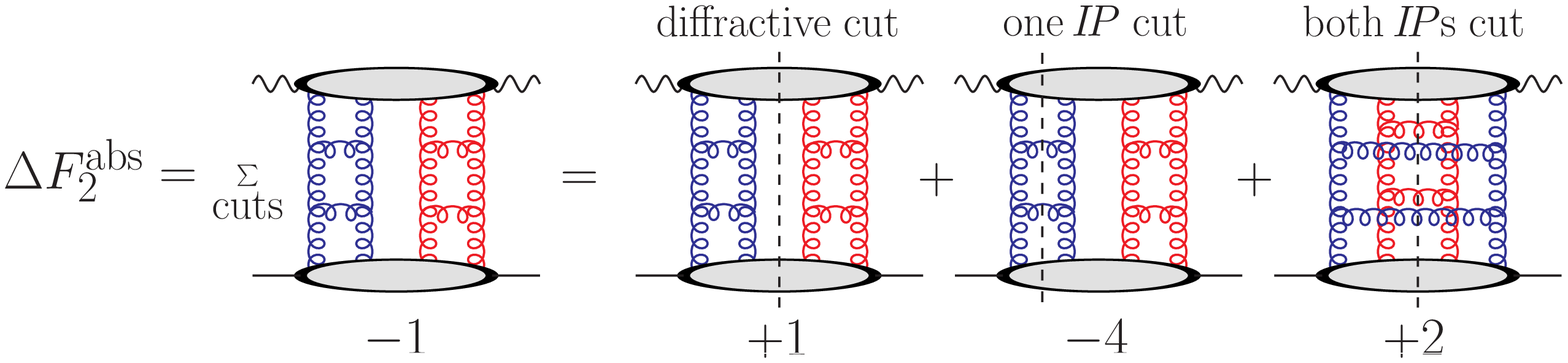}
  \end{minipage}
  \caption{(a) Absorptive corrections to $F_2$ due to the $2\to1$ Pomeron contribution.  (b) Application of the AGK cutting rules.  For simplicity, the upper parton ladder, shown in the right-hand diagram of (a), is hidden inside the upper blob in each diagram of (b).}
  \label{fig:agk}
\end{figure}

The inclusive proton structure function, $F_2(\xB,Q^2)$, as measured by experiment, can be approximately written as a sum of the single Pomeron exchange (DGLAP) contribution and absorptive corrections due to a $2\to 1$ Pomeron merging; see Fig.~\ref{fig:agk}(a).  That is,
\begin{equation} \label{eq:f2total}
  F_2(\xB,Q^2) = F_2^{\mathrm{DGLAP}}(\xB,Q^2) + \Delta F_2^{\mathrm{abs}}(\xB,Q^2).
\end{equation}
In computing $\Delta F_2^{\mathrm{abs}}$ we need to sum over all possible cuts.  The Abramovsky-Gribov-Kancheli (AGK) cutting rules \cite{Abramovsky:1973fm} were originally formulated in Reggeon field theory but have been shown to also hold in pQCD \cite{Bartels:1996hw}.  Application of the AGK rules gives the result that relative contributions of $+1$, $-4$, and $+2$ are obtained according to whether neither Pomeron, one Pomeron, or both Pomerons are cut; see Fig.~\ref{fig:agk}(b).  Therefore, the sum over cuts is equal to \emph{minus} the diffractive cut and so the absorptive corrections can be computed from a calculation of the $t$-integrated diffractive structure function $F_2^{{\rm D}(3)}(\xPom,\beta,Q^2)$, where $\beta\equiv \xB/\xPom$ and $\xPom$ is the fraction of the proton's momentum transferred through the rapidity gap.

The pQCD description of $F_2^{{\rm D}(3)}$ is described in \cite{Martin:2004xw,Martin:2005hd}.  Working in the fixed flavour number scheme (FFNS), it can be written as
\begin{equation}
  \label{eq:F2D3sum}
F_2^{{\rm D}(3)} = \underbrace{F_{2,{\rm non-pert.}}^{{\rm D}(3)}}_{\text{soft Pomeron}} + \underbrace{F_{2,{\rm pert.}}^{{\rm D}(3)} + F_{2,{\rm direct}}^{{\rm D}(3),c\bar{c}} + F_{L,{\rm tw.4}}^{{\rm D}(3)}}_{\text{QCD Pomeron}},
\end{equation}
apart from the secondary Reggeon contribution.  The separation between the soft Pomeron and QCD Pomeron is provided by a scale $\mu_0\sim 1$ GeV.  For simplicity, we take $\mu_0$ to be the same as the scale $Q_0$ at which the input PDFs are taken in the analysis of $F_2$ data, so $\mu_0=Q_0=1$ GeV, the value used in the MRST2001 NLO analysis \cite{Martin:2001es}.  The contribution to the absorptive corrections arising from the soft Pomeron contribution of \eqref{eq:F2D3sum} is already included in the input PDFs, therefore
\begin{equation} \label{eq:deltaF2abs}
  \Delta F_2^\mathrm{abs} = -\frac{1}{1-f_{\text{p.diss.}}}\,\int_{\xB}^1\!\dif{\xPom}\;\left[F_{2,{\rm pert.}}^{{\rm D}(3)} + F_{2,{\rm direct}}^{{\rm D}(3),c\bar{c}} + F_{L,{\rm tw.4}}^{{\rm D}(3)}\right],
\end{equation}
where $f_{\text{p.diss.}}$ is the fraction of diffractive events in which the proton dissociates.  In practice, we take $f_{\text{p.diss.}}=0.5$ and take an upper limit of 0.1 instead of 1 for $\xPom$ in \eqref{eq:deltaF2abs}.\footnote{The value of $f_{\text{p.diss.}}=0.5$ is justified by a ZEUS comparison \cite{Chekanov:2002bz} of proton-tagged diffractive DIS data with data which allowed proton dissociation up to masses of $6$ GeV, where $f_{\text{p.diss.}}=0.46\pm0.11$ was obtained.}

First consider the contribution to \eqref{eq:deltaF2abs} from the $F_{2,{\rm pert.}}^{{\rm D}(3)}$ term.\footnote{The other two contributions to \eqref{eq:deltaF2abs} are described after \eqref{eq:f2global}.}  It corresponds to a $2\to 1$ Pomeron merging with a cut between the two Pomeron ladders and can be written as
\begin{equation} \label{eq:collfact}
  F_{2,{\rm pert.}}^{{\rm D}(3)}(\xPom,\beta,Q^2) = \sum_{a=q,g} C_{2,a}\otimes a_{\rm pert.}^{\rm D},
\end{equation}
where $C_{2,a}$ are the \emph{same} coefficient functions as in inclusive DIS.  The diffractive PDFs, $a^{\rm D}=z q^{\rm D}$ or $z g^{\rm D}$, where $z\equiv x/\xPom$, satisfy an \emph{inhomogeneous} evolution equation \cite{Martin:2005hd}:
\begin{align}
  a_{\rm pert.}^{\rm D}(\xPom,z,Q^2) &= \int_{\mu_0^2}^{Q^2}\!\diff{\mu^2}\;f_{\Pom}(\xPom;\mu^2)\;a^\Pom(z,Q^2;\mu^2) \label{eq:apert} \\
  \Longrightarrow \frac{\partial a_{\rm pert.}^{\rm D}}{\partial \ln Q^2}
  &= \frac{\alpha_S}{2\pi}\sum_{a^\prime=q,g}P_{aa^\prime}\otimes a_{\rm pert.}^{\prime\,{\rm D}}~+~P_{a\Pom}(z)\,f_\Pom(\xPom;Q^2). \label{eq:ddisevol}
\end{align}
Here, $f_\Pom(\xPom;Q^2)$ is the perturbative Pomeron flux factor,
\begin{equation} \label{eq:fPomG}
  f_{\Pom}(\xPom;\mu^2) = \frac{1}{\xPom B_D} \left[\,R_g\frac{\alpha_S(\mu^2)}{\mu}\;\xPom g(\xPom,\mu^2)\,\right]^2.
\end{equation}
The diffractive slope parameter $B_D$ comes from the $t$-integration, while the factor $R_g$ accounts for the skewedness of the proton gluon distribution \cite{Shuvaev:1999ce}.  There are similar contributions from (light) sea quarks, where $g$ in \eqref{eq:fPomG} is replaced by $S\equiv 2(\bar{u}+\bar{d}+\bar{s})$, together with an interference term.  A sum over all three contributions is implied in \eqref{eq:apert} and in the second term of \eqref{eq:ddisevol}.  The Pomeron PDFs in \eqref{eq:apert}, $a^\Pom(z,Q^2;\mu^2)$, are evolved using NLO DGLAP from a starting scale $\mu^2$ up to $Q^2$, taking the input distributions to be LO Pomeron-to-parton splitting functions, $a^\Pom(z,\mu^2;\mu^2)=P_{a\Pom}(z)$ \cite{Martin:2005hd}.

From \eqref{eq:f2total},
\begin{equation} \label{eq:atotal}
  a(x,Q^2) = a^{\mathrm{DGLAP}}(x,Q^2) + \Delta a^\mathrm{abs}(x,Q^2),
\end{equation}
where $a(x,Q^2) = xg(x,Q^2)$ or $xS(x,Q^2)$, and
\begin{equation} \label{eq:deltaaabs}
  \Delta a^\mathrm{abs}(x,Q^2) = - \frac{1}{1-f_{\text{p.diss.}}}\,\int_{x}^1\!\dif{\xPom}\;a_{\rm pert.}^{\rm D}(\xPom,x/\xPom,Q^2).
\end{equation}
Differentiating \eqref{eq:atotal} with respect to $Q^2$ gives the evolution equations for the (inclusive) gluon and sea-quark PDFs:
\begin{equation} \label{eq:disevol}
  \boxed{\frac{\partial a(x,Q^2)}{\partial \ln Q^2}
  = \frac{\alpha_S}{2\pi}\sum_{a^\prime=q,g}P_{aa^\prime}\otimes a^{\prime}~-~\frac{1}{1-f_{\text{p.diss.}}}\,\int_{x}^1\!\dif{\xPom}\;P_{a\Pom}(x/\xPom)\,f_\Pom(\xPom;Q^2).}
\end{equation}
Thus \eqref{eq:disevol} is a more precise version of the GLRMQ equations \eqref{eq:GLRMQgluon}, which goes beyond the DLLA and accounts for sea-quark recombination as well as gluon recombination.  Consider the recombination of gluons into gluons, for example, in the DLLA where $x\ll \xPom$, then $P_{g\Pom}=9/16$ \cite{Martin:2005hd}.  Taking $R_g=1$ and $f_{\text{p.diss.}}=0$, then \eqref{eq:disevol} becomes
\begin{equation}
  \frac{\partial xg(x,Q^2)}{\partial \ln Q^2}
  = \frac{\alpha_S}{2\pi}\sum_{a^\prime=q,g}P_{ga^\prime}\otimes a^{\prime}~-~\frac{9}{16}\,\frac{\alpha_S^2(Q^2)}{B_D\,Q^2}\,\int_{x}^1\!\diff{\xPom}\left[\xPom g(\xPom,Q^2)\right]^2.
\end{equation}
Comparing to \eqref{eq:GLRMQgluon} this is simply the GLRMQ equation with $R^2=8B_D$.  For numerical results we take $B_D = 6$ ($4$) GeV$^{-2}$ for light (charm) quarks, which would correspond to $R = \sqrt{8B_D} = 1.4$ (1.1) fm.

The procedure for incorporating absorptive corrections into a (NLO) global parton analysis (in the FFNS) is as follows:
\begin{enumerate}
  \item Parameterise the $x$ dependence of the input PDFs at a scale $Q_0\sim 1$ GeV.
  \item Evolve the PDFs $xg(x,Q^2)$ and $xS(x,Q^2)$ using the non-linear evolution equation \eqref{eq:disevol}.  (The non-singlet distributions are evolved using the usual linear DGLAP equations.)
  \item Compute
    \begin{equation} \label{eq:f2global}
      F_2(\xB,Q^2) = \sum_{a=q,g} C_{2,a}\otimes a ~-~ \frac{1}{1-f_{\text{p.diss.}}}\,\int_{\xB}^1\!\dif{\xPom}\;\left[F_{2,{\rm direct}}^{{\rm D}(3),c\bar{c}} + F_{L,{\rm tw.4}}^{{\rm D}(3)}\right],
    \end{equation}
    and compare to data.  Here, the two terms inside the square brackets are beyond collinear factorisation, that is, they cannot be written as a convolution of coefficient functions with the PDFs.  The first term inside the square brackets corresponds to the process $\gamma^*\Pom\to c\bar{c}$.  The second term corresponds to the process $\gamma^*\Pom\to q\bar{q}$, for light quarks with a longitudinally polarised photon.  These contributions are calculated as described in Ref.~\cite{Martin:2005hd}.
\end{enumerate}
As usual, these three steps should be repeated with the parameters of the input PDFs adjusted until an optimal fit is obtained.  This procedure is our recommended way of accounting for absorptive corrections in a global parton analysis.  However, in practice, available NLO DGLAP evolution codes, such as the \textsc{qcdnum} \cite{QCDNUM} program, are often regarded as a `black box', and it is not trivial to modify the usual linear DGLAP evolution to the non-linear evolution of \eqref{eq:disevol}.  Therefore, we adopt an alternative iterative procedure which avoids the explicit implementation of non-linear evolution, but which is equivalent to the above procedure.

\section{Effect of absorptive corrections on inclusive PDFs}

We model our analysis of HERA $F_2$ data \cite{f2data} on the MRST2001 NLO analysis \cite{Martin:2001es}, which was the first in which a negative gluon distribution was required at the input scale of $Q_0=1$ GeV.  (The more recent MRST sets have not changed substantially at small $x$.)  We apply cuts $\xB\le0.01$, $Q^2\ge2$ GeV$^2$, and $W^2\ge12.5$ GeV$^2$, leaving $280$ data points.  The input gluon and sea-quark distributions are taken to be
\begin{align}
  \label{eq:inputg}
  xg(x,Q_0^2)~&=~A_g\,x^{-\lambda_g}(1-x)^{3.70}(1+\epsilon_g \sqrt x+\gamma_gx)~-~A_-\,x^{-\delta_-}(1-x)^{10},\\
  \label{eq:inputS}
  xS(x,Q_0^2)~&=~A_S\,x^{-\lambda_S}(1-x)^{7.10}(1+\epsilon_S \sqrt x+\gamma_Sx),
\end{align}
where the powers of the $(1-x)$ factors are taken from \cite{Martin:2001es}, together with the valence-quark distributions, $u_V$ and $d_V$, and $\Delta\equiv \bar{d}-\bar{u}$.  The $A_g$ parameter is fixed by the momentum sum rule, while the other nine parameters are allowed to go free.  Since we do not fit to DIS data with $\xB>0.01$, we constrain the input gluon and sea-quark distributions, and their derivatives with respect to $x$, to agree with the MRST2001 NLO parton set \cite{Martin:2001es} at $x=0.2$.  This is done by including the value of these MRST PDFs at $x=0.2$, and their derivatives, as data points in the fit, with an error of 10\% on both the value of the MRST PDFs and their derivatives.  Therefore, the PDFs we obtain are not precisely constrained at large $x$, but this paper is primarily concerned with the small-$x$ behaviour of the PDFs.

The procedure we adopt is as follows:
\begin{enumerate}
  \renewcommand{\labelenumi}{(\roman{enumi})}
\item Start by performing a standard NLO DGLAP fit to $F_2$ data with no absorptive corrections.
\item Tabulate $\Delta F_2^{\mathrm{abs}}$, given by \eqref{eq:deltaF2abs}, and $\Delta a^{\mathrm{abs}}$, given by \eqref{eq:deltaaabs}, using PDFs $g(\xPom,\mu^2)$ and $S(\xPom,\mu^2)$ obtained from the previous fit.
\item Perform a standard NLO DGLAP fit to `corrected' data, $F_2^{\mathrm{DGLAP}}=F_2-\Delta F_2^{\mathrm{abs}}$, to obtain PDFs $a^{\mathrm{DGLAP}}$.  Then correct these PDFs to obtain $a = a^{\mathrm{DGLAP}} + \Delta a^\mathrm{abs}$.  These latter PDFs $a$ then satisfy the non-linear evolution equations \eqref{eq:disevol}.
  \item Go to (ii).
\end{enumerate}
Each successive iteration of steps (ii) and (iii) introduces another level of $2\to 1$ Pomeron mergings, so that eventually all the `fan' diagrams are included, achieving the same effect as the procedure described at the end of Section \ref{sec:non-linear}.

Note that the correction to the PDFs, $a = a^{\mathrm{DGLAP}} + \Delta a^\mathrm{abs}$, in each step (iii), was omitted in our previous analysis \cite{Martin:2004xx}.  Consequently, the effect of the absorptive corrections on the PDFs at large scales was overestimated.  Also in \cite{Martin:2004xx}, the known LO $P_{a\Pom}(z)$ were multiplied by free parameters (`K-factors'), determined from separate fits to diffractive DIS data, in an attempt to account for higher-order pQCD corrections to the LO Pomeron-to-parton splitting functions.  However, since these K-factors took unreasonable values, with some going to zero, here we have chosen to fix them to 1.  Therefore, the updated analysis, presented here, does not require a simultaneous fit to the diffractive DIS data.

\begin{figure}
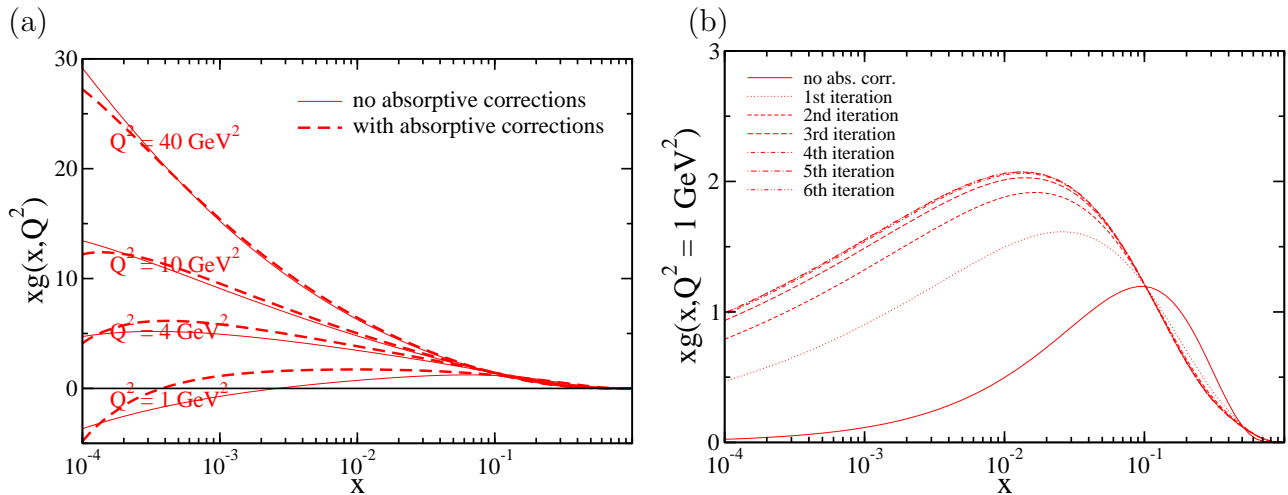

  \centering
  \begin{minipage}{\textwidth}
    (a)\hspace{0.5\textwidth}(b)\\
    \includegraphics[width=0.49\textwidth,clip]{neggluon.eps}\hfill
    \includegraphics[width=0.49\textwidth,clip]{posgluon.eps}
  \end{minipage}
  \caption{(a) The gluon distribution obtained from fits to $F_2$ data, before and after absorptive corrections have been included.  (b) The effect of successive iterations on the gluon distribution obtained from fits to $F_2$, taking a positive definite input gluon at 1 GeV.  Each iteration introduces another level of $2\to 1$ Pomeron mergings.}
  \label{fig:gluon}
\end{figure}

In Fig.~\ref{fig:gluon}(a) we show the gluon distribution at scales $Q^2 = 1$, $4$, $10$, and $40$ GeV$^2$ obtained from fits before and after absorptive corrections have been included.  Both fits are almost equally good with $\chisq$ values of $0.86$ and $0.87$ for the fits without and with absorptive corrections respectively.  At low $Q^2$ the absorptive corrections give an increased gluon distribution at small $x$, apart from at $x\lesssim 10^{-4}$ where there are only a few data points and where additional absorptive effects (Pomeron loops) may become important.  The non-linear term of \eqref{eq:disevol} slows down the evolution, so that by $40$ GeV$^2$ the two gluon distributions are roughly equal; see Fig.~\ref{fig:gluon}(a).

We repeated the fits without the negative term in the input gluon distribution, that is, without the second term in \eqref{eq:inputg}.  When absorptive corrections were included, almost the same quality of fit was obtained ($\chisq = 0.90$), while without absorptive corrections the fit was slightly worse ($\chisq = 0.95$).  We conclude that absorptive corrections lessen the need for a negative gluon distribution at $Q^2 = 1$ GeV$^2$.  The gluon distributions obtained from six successive iterations of steps (ii) and (iii) above are shown in Fig.~\ref{fig:gluon}(b).  The convergence is fairly rapid, with only the first three iterations having a significant effect, that is, the `fan' diagrams which include $8\to 4\to 2\to 1$ Pomeron mergings.

Although we have seen that the inclusion of absorptive corrections has reduced the need for a \emph{negative} gluon, it has not solved the problem of the \emph{valence-like} gluon.  That is, the gluon distribution at low scales still decreases with decreasing $x$, whereas from Regge theory it is expected to behave as $xg\sim x^{-\lambda_{\text{soft}}}$ with $\lambda_{\text{soft}}\simeq 0.08$.  We have studied several possibilities of obtaining a satisfactory fit with this behaviour \cite{Martin:2004xx}.  The only modification which appears consistent with the data (and with the desired $\lambda_g = \lambda_S$ equality) is the inclusion of power-like corrections, specifically, a global shift in all scales by about 1 GeV$^2$.  (Note that a similar shift in the scale is required in the dipole saturation model \cite{dipole}.)  However, we do not have a solid theoretical justification for this shift.  Therefore, a more detailed, and more theoretically-motivated, investigation of the effect of power corrections in DIS is called for.

\section*{Acknowledgements}

We thank Robert Thorne for useful discussions.  ADM thanks the Leverhulme Trust for an Emeritus Fellowship.  This work was supported by the UK Particle Physics and Astronomy Research Council, by the Federal Program of the Russian Ministry of Industry, Science and Technology (grant SS-1124.2003.2), by the Russian Fund for Fundamental Research (grant 04-02-16073), and by a Royal Society Joint Project Grant with the former Soviet Union.

\end{document}